\documentclass{elsart}
\usepackage{epsfig,graphicx, amssymb}
\begin{document}
\runauthor{Borowiec et al.}

\begin{frontmatter}
\title{
Response of a Magneto-Rheological Fluid Damper Subjected to Periodic Forcing in 
a High Frequency Limit
}
\author[Lublin]{Marek  Borowiec}
\author[Lublin,Wien]{, Grzegorz Litak\thanksref{E-mail}}
\author[Lublin]{and Rafa\l{} Kasperek}

\address[Lublin]{Department of Applied Mechanics, Technical University of
Lublin, 
Nadbystrzycka~36, PL-20-618 Lublin, Poland}
\address[Wien]{Institut f\"{u}r Mechanik und Mechatronik, Technische 
Universit\"{a}t 
Wien,
 Wiedner
Hauptstra{\ss}e
8 - 10, A-1040 Wien, Austria} 

\thanks[E-mail]{Fax: +48-815250808; E-mail:
g.litak@pollub.pl (G. Litak)}

\begin{abstract}
We explored vibrations of a single-degree of freedom oscillator
with a magneto-rheo\-lo\-gi\-cal  damper subjected to kinematic excitations. 
Using fast and
slow scales decoupling procedure we derived an effective damping coefficient in the limit of high 
frequency excitation. 
Damping characteristics, as functions of  velocity, change considerably especially by terminating the 
singular non-smoothness points.
This effect  was more transparent for a larger control parameter which was defined as 
the product of the 
excitation amplitude and its frequency.  
\end{abstract}
\begin{keyword}
mgneto-rheological fluid damper, nonlinear vibrations, multiple time-scales
\end{keyword}

\end{frontmatter}
{\em PACS}: 75.80.+q, 83.80.Gv, 05.45.-a, 46.40.-f


\section{Introduction}
Magneto-rheological (MR) semi-active dampers 
with  multiple ratio 
characteristics 
are able to be controlled efficiently by  applied magnetic field vibrations 
of various dynamical systems \cite{Spencer1997,Savaresi2005,Borowiec2006,Guo2006}.
Having relatively small switching time (between different damping ratio) they have 
replaced traditional electro-hydraulic dampers \cite{Fisher2004}.
In this paper we will focus on an effective damping term in the presence of fast kinematic 
oscillations. To describe the  damping phenomenon of the suspension quarter-car model, 
we use the Bingham viscoplastic model \cite{Spencer1997,Savaresi2005}, 
which assumes a mixture of viscous  
and dry friction terms (composed of piece linear regions). To obtain the effective damping term we will use the two time scales
splitting which is relevant in large frequency kinematic excitation due 
to the motion of car on the rough
road surface.

The present paper in divided into four sections.
Following the Introduction (Sec. 1) we will present the Model (Sec. 2).
The calculation schema and the results obtained will be offered in Sec. 3,
Calculations and Results.
We will  finish with Sec. 4, Summary and Conclusions. 
 
\begin{figure}[htb]
 \centerline{
 \epsfig{file=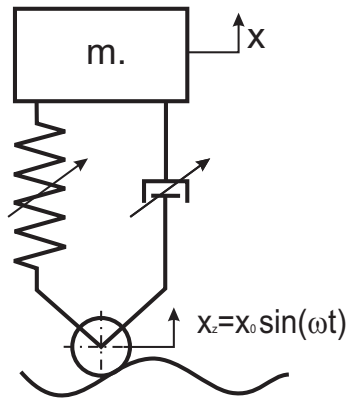,width=4.5cm,angle=0}}
 \caption{ 
   \label{fig1}
 A single degree of freedom of quarter--car model.}
\end{figure}  

\begin{figure}[htb]
 \centerline{
\includegraphics[width=6cm,angle=-90]{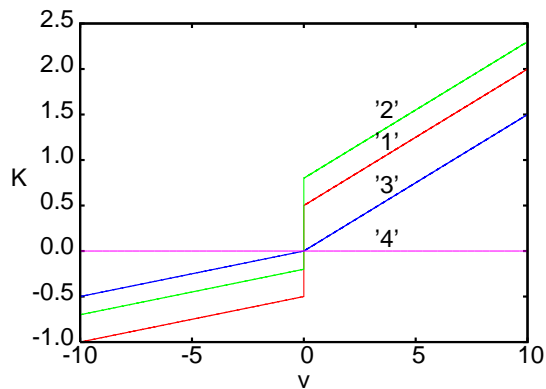}}
 \caption{
   \label{fig2}
Damping term $K$ against velocity for $\alpha=0.5$ and 
$k_1=0.1$,  $k_2=0.5$, $\Delta k=0.0$ for '1';
$k_1=0.1$,  $k_2=0.5$, $\Delta k=0.3$ for '2';
 $k_1=0.1$,  $k_2=0.0$, $\Delta k=0.0$ for '3';
$k_1=0.0$,  $k_2=0.0$, $\Delta k=0.0$ for '4'.}
\end{figure}


\section{The model}
The dimensionless equation of motion for a single degree of freedom
oscillator frequently used in description of a quarter-car model \cite{VonWagner2004}
\begin{equation}
\label{eq1}
\ddot x +    K(\dot x) + x= \omega^2 x_0\sin \omega t,
\end{equation}
where  $-K(\dot x)$ is a damping force and $\omega^2 x_0\sin \omega t$ is a kinematic excitation term
dependent on the vehicle speed and the roughness of the road surface. For simplicity we assumed that this surface can be
described by a harmonic function (Eq. \ref{eq1}).

\begin{figure}[htb]
 \centerline{
\includegraphics[width=6cm,angle=-90]{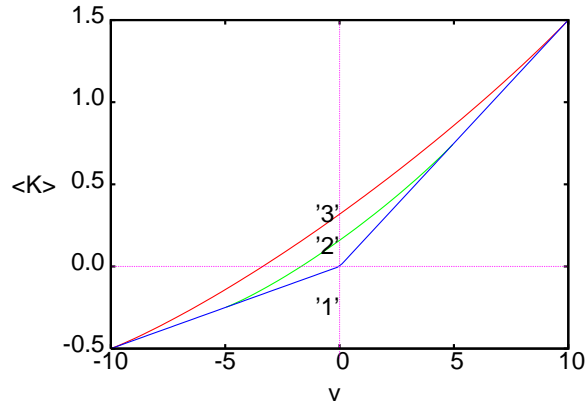}}
 \caption{ 
   \label{fig3} Averaged damping term $<K>$ against velocity $v$
where $k_1=0.1$,  $k_2=0.5$, $\Delta k=0.0$ for '1';
$k_1=0.1$,  $k_2=0.5$, $\Delta k=0.3$ for '2';
 $k_1=0.1$,  $k_2=0.0$, $\Delta k=0.0$ for '3'(as in Fig. \ref{fig1}), and
 $\omega x_0=10$.}
\end{figure}

Here we assumed that the nonlinear damping force $-K(\dot x)$  is of  Bingham type 
\cite{Spencer1997,Savaresi2005}

\begin{equation}
\label{eq2}
K(\dot x)= k_1(1 + \alpha {\rm sgn}\dot x ) \dot x +k_2 {\rm sgn}\dot x
 + \Delta k.
\end{equation} 

Some examples of possible damping characteristics are presented in Fig. \ref{fig2}.
Note, the above force term possesses singular points where the damping is non-continuous (lines '1','2') or non-smooth (line '3').
Usually, the mirror symmetry  ($v$ $\rightarrow$ $-v$) is broken by the change in slopes (line '3') \cite{VonWagner2004}. 
Additionally, it can be also associated with constant term ('2' and '3') related to the direction dependent 
dry friction phenomenon \cite{Thomsen2003}.   
Using the above model we will investigate the effect of oscillations with high frequency. For a quarter-car model  
(Fig.  \ref{fig1})
this effect could appear during its motion with high speed.

\section{Calculations and Results}

The large value of excitation frequency $\omega >> 1$ introduces naturally
to second 'fast' time scale in our system (Fig. \ref{fig1}).
     
In fact the multiple time-scale method  is an efficient and widely used treatment for nonlinear mechanical systems
 \cite{Nayfeh1979,Cartmell2003,Warminski2003}
to get an approximate analytic solution. In these cases the time-scales are introduced to account for nonlinear terms
in the examined systems.
On the other hand, the fast time scale can be introduced through excitation terms in the limit of very high
excitation frequency \cite{Thomsen2003,Thomsen2005,Chatterjee2003,Litak2006}.
In various nonlinear systems  \cite{Thomsen2003}, subjected to
additional excitation  fast and slow time-scales, such  splitting is realized physically.
It enables to
estimate a behaviour in terms of  the averaged system equations of motion after averaging over fast oscillations.

Note, comparing to our system (Eqs. \ref{eq1},\ref{eq2}) in the original treatment by Thomsen, $k_1=0$ $k_2 \neq 0$ and $\Delta 
k 
\neq 0$ \cite{Thomsen2003,Thomsen2005} to
describe the effect of dry friction.
On the other hand, in Ref. \cite{VonWagner2004} $k_2=0$ and $\Delta k=0$, $k_1 \neq 0$ for a quarter-car model.
Below we will consider the most general case with damping (Eq. \ref{eq2})
described by all nonzero terms.

Assuming that $\omega >>  1$  we define second (fast) time scale by introducing a small parameter:

\begin{equation}
\label{eq3}
\epsilon= \omega^{-1}
\end{equation}
 Following  Thomsen (2005) \cite{Thomsen2005}
we define
 two time scales fast given by
$T_0$ and  slow by $T_1$, respectively

\begin{equation}
\label{eq4}
T_0= \epsilon^{-1} t,~~~~T_1=t.
\end{equation}
Then
two corresponding time derivatives $D_0$ and $D_1$ are

\begin{equation}
\label{eq5}
D_0= \partial/ \partial T_0,~~~~~D_1= \partial/ \partial T_1.
\end{equation}

The original time derivatives can be written as follows
\begin{eqnarray}
\label{eq6} 
&& \frac{\rm d}{{\rm d} t} = \epsilon^{-1} D_0 + D_1, \\
&& \frac{{\rm d}^2}{{\rm d} t^2} = \epsilon^{-2} D_0^2 + 2 \epsilon^{-1}  D_0 D_1 + D_1^2, \nonumber
\end{eqnarray}
while
a two component solution reads
\begin{equation}
\label{eq7}
x(T_0,T_1)=z(T_1)+ \epsilon \xi(T_0,T_1).
\end{equation}

By introducing the above quantities (Eqs. \ref{eq4},\ref{eq7}) into the initial equation of motion
(Eq. \ref{eq1}) and  collecting terms by  
$\epsilon^{-1}$
we obtain equation for a fast motion part
\begin{equation}
\label{eq8}
D_0^2 \xi = \omega x_0 \sin(T_0).
\end{equation}

On the other hand  for a slow motion one gets (collecting terms by
$\epsilon^0$) 

\begin{equation}
\label{eq9}
D_1^2z + K(D_1 z+ D_0 \xi )+ z =0.
\end{equation}

To find a solution for $z$ in  the last equation (Eq. \ref{eq9}) we have
to
examine fast oscillations in $\xi$ and perform averaging of  the whole
equation
in the excitation period $2 \pi $ (in the fast $T_0$ time  scale).
   
At first, we get a solution for $\xi$  from Eq. \ref{eq8}

\begin{figure}[htb]
 \centerline{ 
\includegraphics[width=5cm,angle=-90]{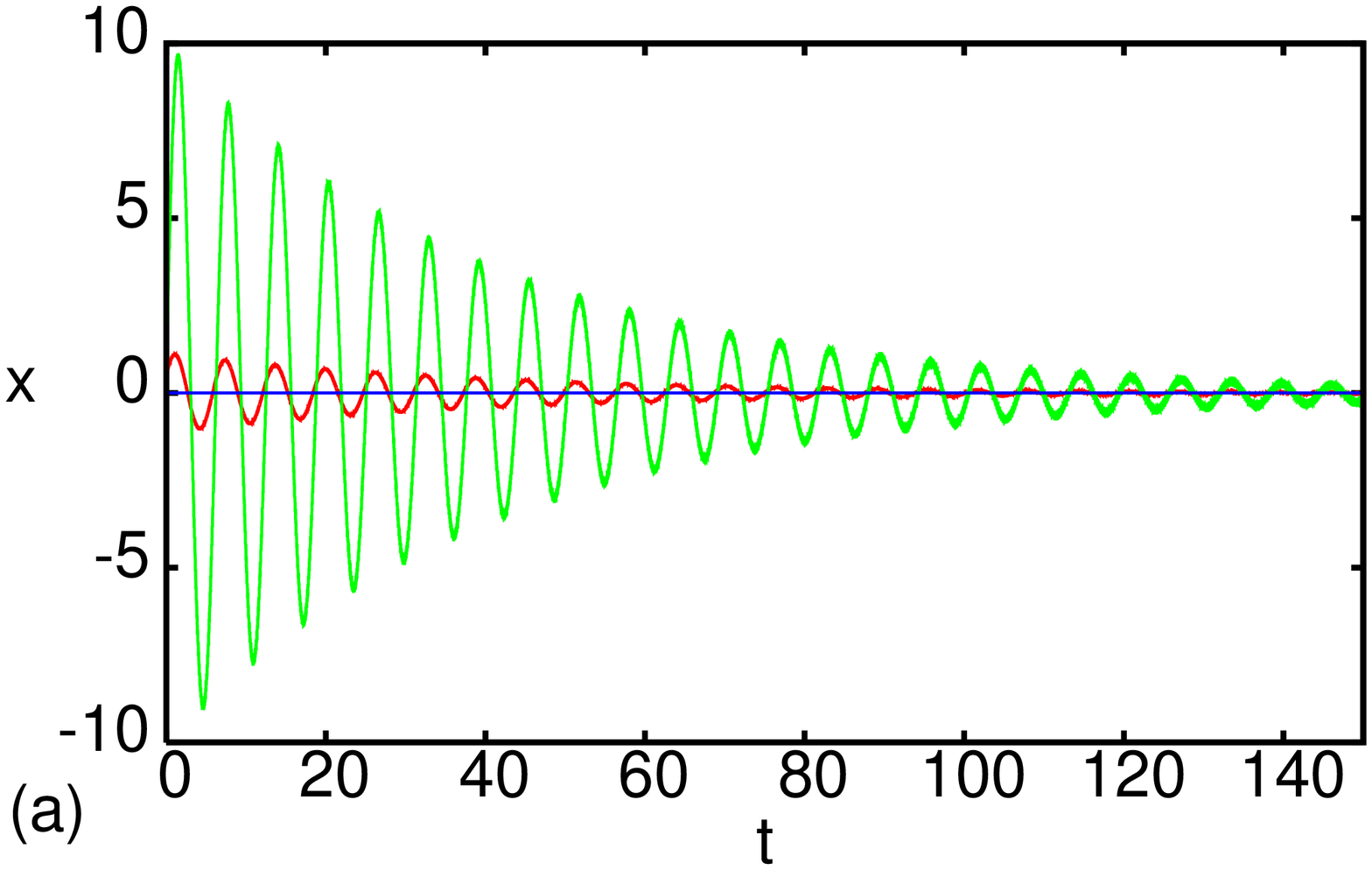} \hspace{-1cm}
\includegraphics[width=5cm,angle=-90]{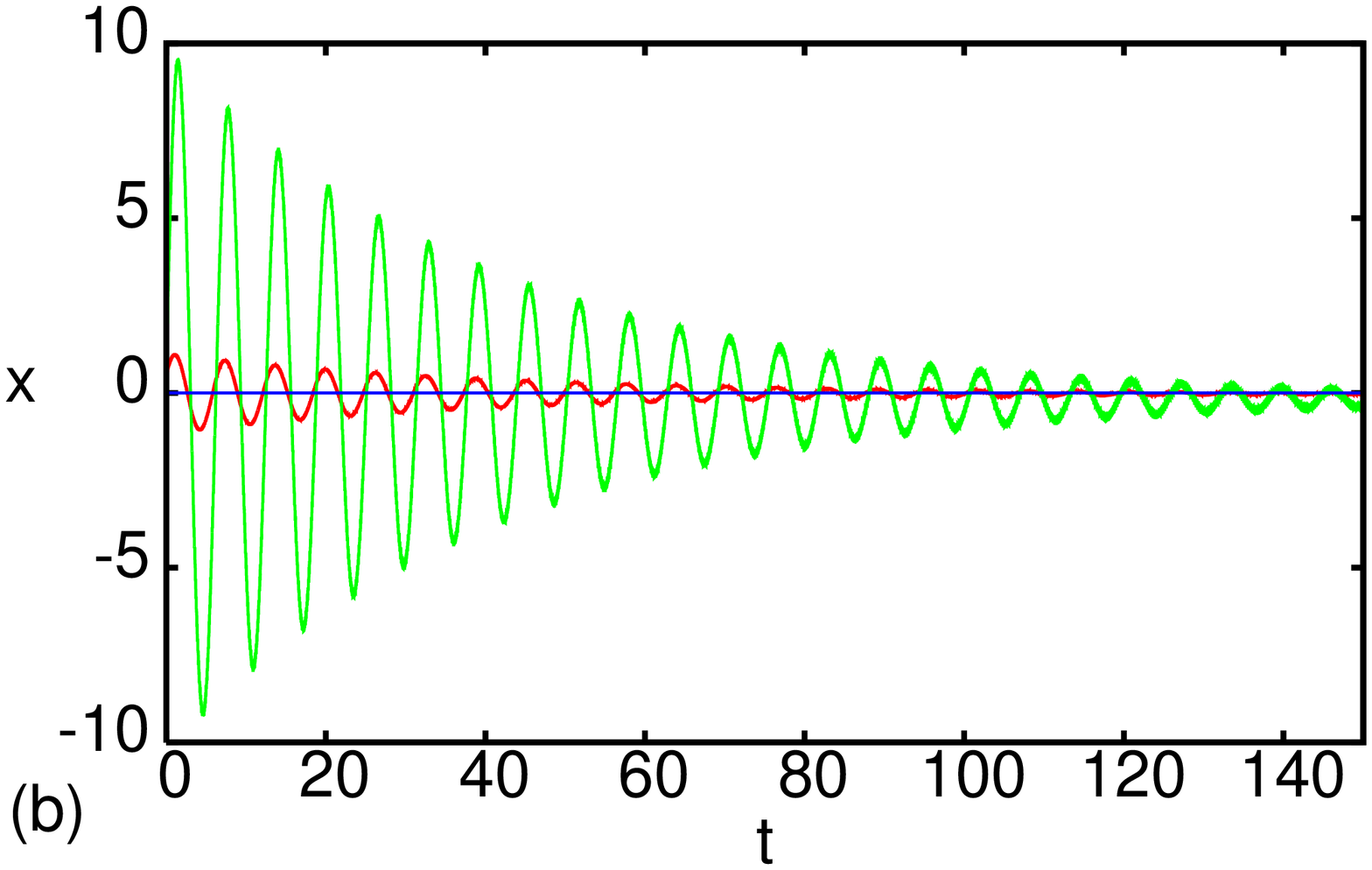}
}
 \centerline{   
\includegraphics[width=5cm,angle=-90]{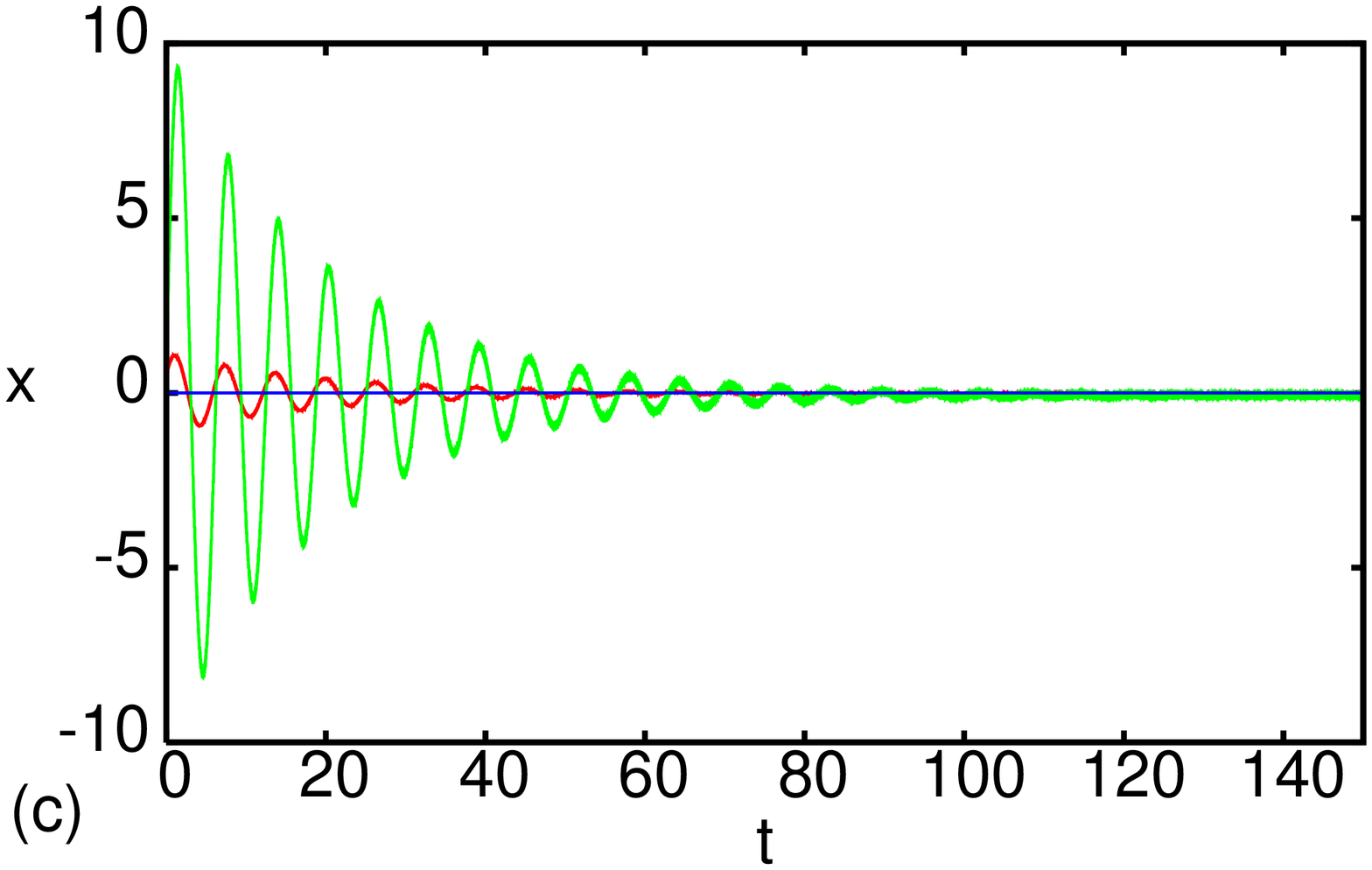} \hspace{-1cm}
\includegraphics[width=5cm,angle=-90]{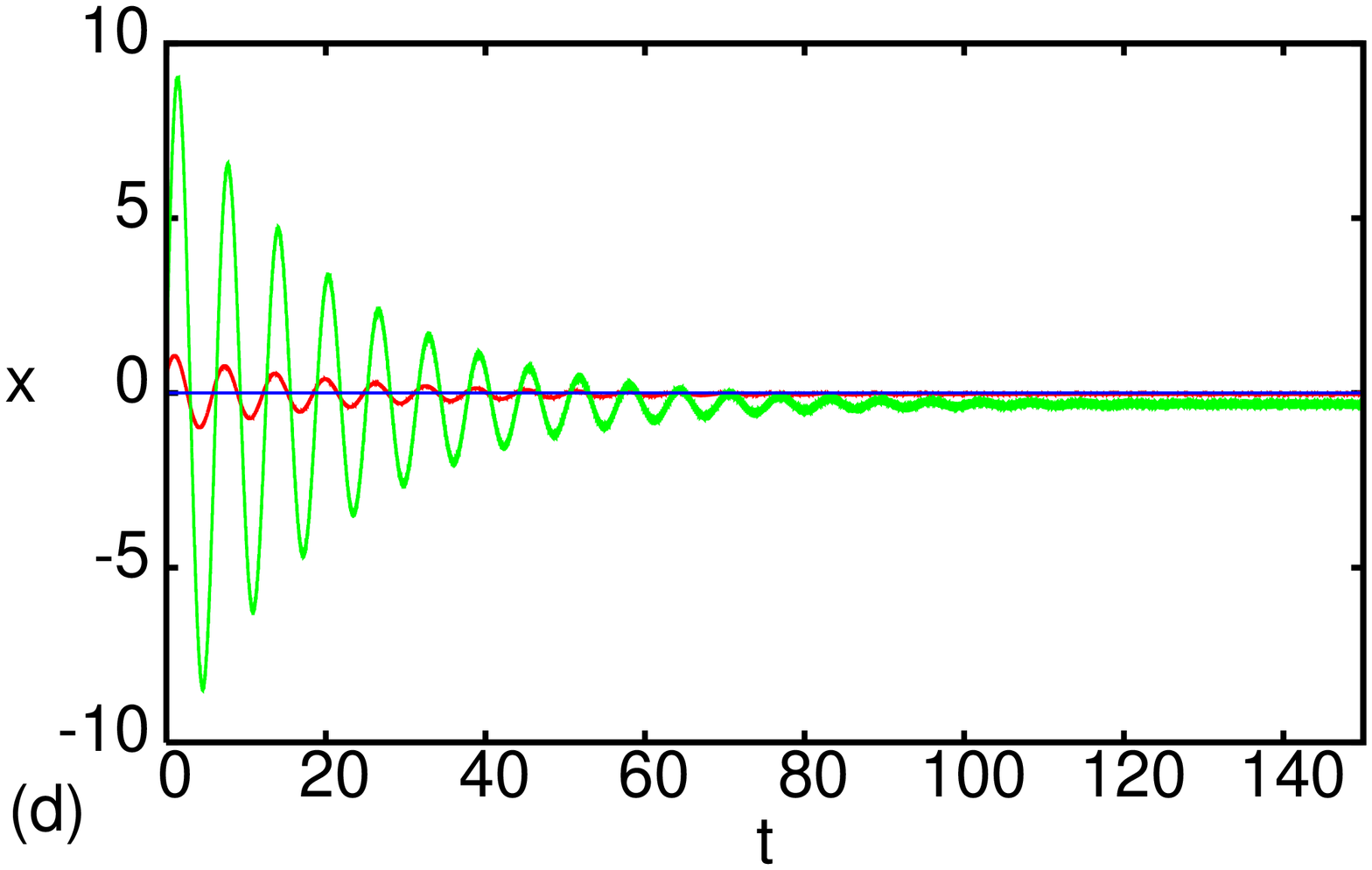}
}
 \caption{ Time histories of damping term for system
parameters: $k_1=0.05$, $k_2=0$, $\Delta k=0$, $\omega=100$ and
(a) $\alpha=0.1$,
(b) $\alpha=0.5$,
(c) $\alpha=0.1$,
(d) $\alpha=0.5$,
and $x_0=0.01$ (red line), $x_0=0.1$ (green line) respectively. \label{fig4}}
\end{figure}

\begin{figure}[htb]
 \centerline{
\includegraphics[width=6cm,angle=-90]{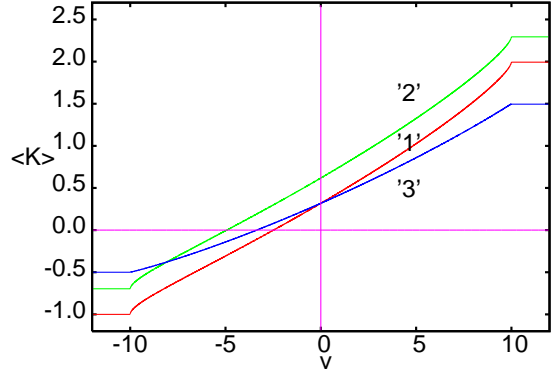}}
 \caption{
   \label{fig5} Averaged values of a damping term $<K>$ against velocity $v$
($\alpha=0.5$, $k_1=0.1$) without excitation '1' and with an excitation term
'2'--'3'
for $\omega x_0=5$ and 10, respectively. Note the curve '3' may correspond
to
case $x_0=0.1$ and $\omega=100$.
}
\end{figure}

\begin{equation}
\label{eq10}
\xi(T_0)= -\omega x_0 \sin(T_0).
\end{equation}

Then averaging procedure for a damping term can be written as

\begin{eqnarray}
\label{eq11}
<K(D_1 z+ D_0 \xi) > &=& 
\frac{1}{2 \pi}\int_0^{2 \pi} \left(
k_1(1 + \alpha {\rm sgn}(D_1 z+ D_0 \xi ))(D_1 z+ D_0 \xi) \right.
\nonumber
\\
&+& \left. k_2 {\rm
sgn} (D_1 z+ D_0 \xi ) ) + \Delta k
\right) {\rm d} \tau
\end{eqnarray}

By using Eq. \ref{eq10} and evaluating the above integral   
we get
\begin{eqnarray}
<K> &=& k_1 \alpha \frac{1}{2 \pi}\int_0^{2 \pi} ( D_1z -x_0 \omega \cos
\tau) {\rm sgn}( D_1z -x_0 \omega \cos
\tau) {\rm d} \tau \nonumber  \\ &+& k_2  \frac{1}{2 \pi}\int_0^{2 \pi}
{\rm sgn}( D_1z -x_0 \omega \cos
\tau) {\rm d} \tau + \Delta k,
\label{eq12}
\end{eqnarray} 
where $\tau= T_0 \in [0,2\pi]$.

After integration we get an analytical formula for the averaged damping term 
\begin{eqnarray}
<K> &=& \left[ 1 -\frac{2}{\pi}\arccos{\left(\frac{ D_1z}{x_0
\omega}\right)} \right] (k_2 +k_1 \alpha
D_1z)
\nonumber \\
&+&
\frac{2}{\pi} \sin \left[\arccos{\left(\frac{ D_1z}{x_0 \omega}\right)} 
\right]  k_1 \alpha
x_0 \omega 
 + k_1 D_1z + \Delta k. \label{eq13}
\end{eqnarray}
It is obvious that $x_0 \omega$ is a control parameter which scaling influences 
 averaged damping in a nontrivial way.

Following Thomsen \cite{Thomsen2003,Thomsen2005} discussion on dry friction, we
decided to plot  the
 results (Eq. \ref{eq13}) for three values of  $x_0 \omega$  and $k_2=0$ and $\Delta k=0$ in Fig. \ref{fig3}.
For comparison 
 we have also performed the simulations of the system, 
using Eq. \ref{eq1},
for large $\omega$ ($\omega=100$). Figures
4a--d show the results for different choice of system parameters.
The simulation results show that the product $x_0 \omega$ and $\alpha$ determine the effective damping.
It is visible directly in  envelops of quenching curves as well as in the asymptotic equilibrium value
$<x>=\lim_{t \rightarrow \infty} x$. In our simulation cases Figs. 4 a--d
we have got $<x> \approx$ 0, -0.02, -0.006, -0.031 for red lines;  $<x> \approx$ 0, -0.2, -0.06, -0.31 for green
lines. Note that the simulation result shown in Fig. 4d  $<x> \approx -0.31$ (for the green line) agrees
with the analytic result (Eq. 12, Fig. \ref{eq3} -- line '3'). This agreement is due to the equilibrium of effective damping
for $v=0$ 
and the static displacement force (note, the stiffness coefficient is equal to 1).
Finally, in Fig. \ref{fig5} we show the effective damping $<K>$ for various choices of damping terms 
(Eq. \ref{eq2}) including asymmetric friction  phenomenon related to curves plotted in Fig. \ref{fig2}.
When comparing these two figures (Figs. \ref{fig2} and \ref{fig5}) one can observe that fast oscillations can easily eliminate the 
non-smoothness and discontinuities in the original characteristics.  As we expected,  asymmetries in the initial damping 
terms (Eq. 
\ref{eq2})
lead to new equilibrium point.

\section{Summary and Conclusions}

We have examined vibrations of a single-degree of freedom oscillator with MR  fluid damper
subjected
to kinematic excitations.  Nontrivial behaviour of the system is immanent feature of  
 the MR fluid damper.
For
defined magnetic fluid characteristics \cite{Savaresi2005,VonWagner2004}, we have used the procedure based on fast and slow 
scales 
splitting and, in this way,
we have derived
effective damping coefficient depending on magnetic fluid characteristics.  Consequently, we have shown how the initial oscillations
were quenched out in presence of fast kinematic forcing. Presence of asymmetry in the original damping terms (Eq. \ref{eq2})
changes the equilibrium point.

We should add that the effect of fast excitations can appear in a different way including external forcing of various  
kinds and an additional degree of freedom.  
Particularly this effect can be used to
improve brake systems in cars \cite{Toon2003}.

\section*{Acknowledgements} This research has been partially supported by the 6th Framework Programme,
Marie Curie Actions, Transfer of Knowledge, Grant No. MTKD-CT-2004-014058. Authors would like to thank prof. 
E.M. Craciun for helpful discussions.

\end{document}